\documentclass[runningheads]{llncs}
\usepackage[T1]{fontenc}
\usepackage{graphicx}
\usepackage{enumitem}
\begin{document}
\title{Applying Large Language Models in Knowledge Graph-based Enterprise Modeling:\\Challenges and Opportunities}
%
\titlerunning{Large Language Models in Knowledge Graph-based Enterprise Modeling}
%
\author{Benedikt Reitemeyer\inst{1}\orcidID{0009-0001-4131-1716} \and
Hans-Georg Fill\inst{1}\orcidID{0000-0001-5076-5341} }
\authorrunning{Reitemeyer and Fill}
%
\institute{Research Group Digitalization and Information Systems, University of Fribourg, Fribourg, Switzerland\\
\email{benedikt.reitemeyer|hans-georg.fill@unifr.ch}\\
\url{http://www.unifr.ch/inf/digits}}
\maketitle              
\begin{abstract}
The role of large language models (LLMs) in enterprise modeling has recently started to shift from academic research to that of industrial applications. Thereby, LLMs represent a further building block for the machine-supported generation of enterprise models. In this paper we employ a knowledge graph-based approach for enterprise modeling and investigate the potential benefits of LLMs in this context. In addition, the findings of an expert survey and ChatGPT-4o-based experiments demonstrate that LLM-based model generations exhibit minimal variability, yet remain constrained to specific tasks, with reliability declining for more intricate tasks. The survey results further suggest that the supervision and intervention of human modeling experts are essential to ensure the accuracy and integrity of the generated models. 


\end{abstract}
\section{Introduction}
The utilization of Large Language Models (LLMs) in enterprise modeling has undergone a significant evolution in recent years, progressing from their initial role as a subject of academic research to their current status as a tool employed in industrial applications. Research has demonstrated the potential of LLM-based approaches, yielding impressive results across a range of use cases~\cite{fill2023conceptual,BarnBS23}. This has led to more complex investigations, e.g.\ including process mining techniques~\cite{berti2023abstractions}. In the industrial context, tools such as ADONIS\footnote{\url{https://www.boc-group.com/en/blog/bpm/adonis-and-ai/}} or SAP Signavio\footnote{\url{https://www.signavio.com/process-ai/}} have introduced LLM-based artificial intelligence assistants to facilitate modeling activities.
Another area of enterprise modeling where the use of LLMs appears promising is enterprise architecture modeling. The complex modeling languages employed in this domain, such as Open Group's ArchiMate, present significant challenges for human modelers. The creation of such models is often time-consuming, involves multiple stakeholders, employs diverse concepts and relationships, and is conducted at a fast pace, resulting in models that may quickly become outdated~\cite{aier2009survival,buschle2012tool}.

One of the key challenges in the field of enterprise modeling lies in the inherent complexity of both the modeling languages and the domains and systems to be modeled. Machine-supported approaches based on 
LLMs have the potential to assist modelers in accelerating the modeling process and improving model quality by suggesting appropriate model elements for the given context. However, before addressing such tasks, it is necessary to investigate more deeply how LLMs can be utilized for modeling tasks and identify their limitations in comparison to human modeling activities.
Therefore, we investigated the performance of \emph{human experts} and \emph{ChatGPT-4o} in mapping ArchiMate modeling elements to domain descriptions. We reverted to a modeling approach using knowledge graphs, which can bring benefits in semantics systems engineering~\cite{BUCHMANN2024}, e.g.\ in terms of interoperability and model processing. Three research questions (RQ) were defined both for \emph{human actors} and the \emph{AI agents}:

\begin{itemize}[itemindent=1em]
\item[RQ1:] In what sequence are ArchiMate elements for a specific viewpoint prioritized in relation to a given domain concept?
\item[RQ2:] What is the probability that an ArchiMate element from a specific viewpoint will be proposed as instance for a given domain concept?
\item[RQ3:] What is the proposed relationship type between ArchiMate elements and a given domain concept?
\end{itemize}

Starting with a brief overview on the foundations of enterprise modeling, semantics and large language models in Section~\ref{sec:foundations}, the paper proceeds with an overview of the various methods through which LLM integration can be achieved in enterprise modeling (Section~\ref{sec:integration}). In Section~\ref{sec:evaluation}, the methodology employed for the expert survey and LLM-based experiments is outlined, along with a detailed account of the findings. The results and their implications on usage scenarios and limitations are further discussed in Section~\ref{sec:discussion}. The paper closes with a brief conclusion and outlook on further research steps in Section~\ref{sec:conclusion}.

\section{Foundations}
\label{sec:foundations}
This section briefly introduces some foundations for our experiments in machine-supported enterprise model generation. This concerns the role of semantic in enterprise modeling, the relationship between large language models and semantics, and finally the current state of using large language models in enterprise modeling. 


\subsection{Enterprise Modeling and Semantics}
\begin{figure}
\includegraphics[width=\textwidth]{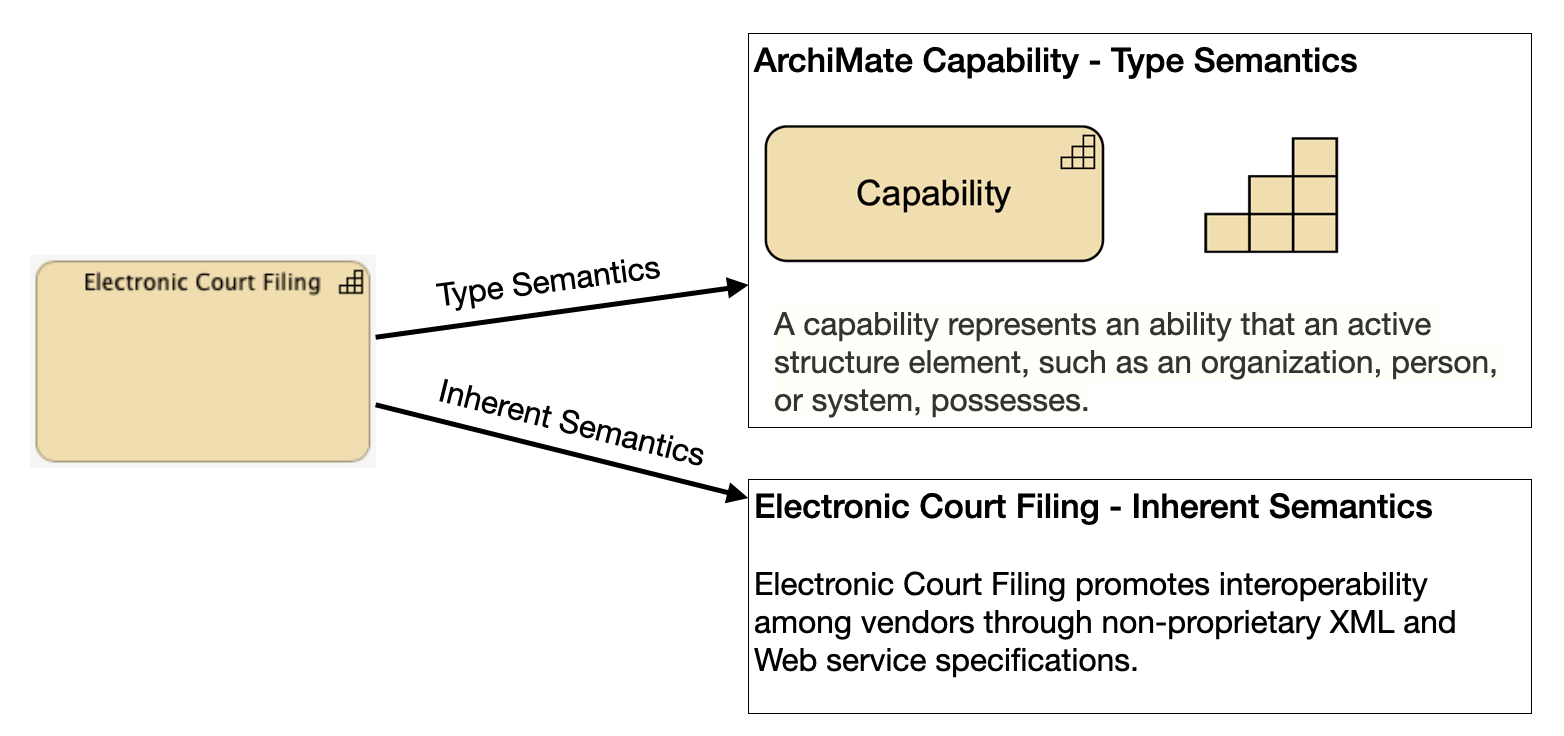}
\caption{Example for type semantics and inherent semantics for an ArchiMate capability element instantiated as Electronic Court Filing.} \label{fig1}
\end{figure}

Modeling the different aspects of enterprises has become a valuable task over the last decades and even more so in the context of digital transformation~\cite{SandkuhlFHKMOSU18}. It aims at visualizing and formally representing enterprise structures and behaviors. Concepts captured in enterprise models include overviews on actors and their roles in the enterprise, business processes, information system landscapes, or application structures~\cite{Vernadat20}.\\
In the scope of this paper, enterprise modeling is seen as a sub-discipline of language-based conceptual modeling that bases on formal syntax in the form of a grammar, i.e.\ symbols and rules for their combination~\cite{VisicFBK15,fill2020enterprise}. On top of syntax and symbols, semantics is defined as a mapping between elements of the grammar and a semantic domain~\cite{HarelR04}.
Semantics can be further classified in \emph{type semantics} and \emph{inherent semantics}~\cite{hofferer2007achieving} -- see Figure~\ref{fig1}\footnote{All Figures and use cases are provided here: https://drive.switch.ch/index.php/s/PvhdETOaAXbNyUN.}. Type semantics refers to the meaning of the elements of the modeling language, e.g. the concept of capabilities in business capability models. Inherent semantics refers to the meaning assigned at the time of instantiation to elements of the modeling language, e.g.\ typically in the form of labels and attribute values.
Lastly, pragmatics relate to the context, goals and purposes, use and users, and effects of a modeling language and the models~\cite{Zemanek66}. 

Due to their capability of formally expressing semantics and making them machine-processable, \emph{ontologies} and \emph{knowledge graphs} (KG) were found to be useful in supporting the automated generation of enterprise models.
Ontologies are characterized as ”a shared and common understanding of some domain that can be communicated across people and computers”~\cite{studer1998knowledge}. Ontologies are represented in formal notation and are based on axioms for enabling reasoning and inferencing to derive new knowledge~\cite{guarino1998formal}. They typically exhibit a high semantic expressiveness using logic-based languages such as the Web Ontology Language (OWL)~\cite{feilmayr2016analysis}. Following the characterization of Ehrlinger and W{\"o}{\ss}, we consider ontologies as formal knowledge bases, which are used by knowledge graphs to "acquire and integrate information into an ontology"~\cite{ehrlinger2016towards}. In addition, they have the capability to derive new knowledge through reasoning and describe relevent real-world entities as provided by open knowledge bases like DBpedia, schema.org or YAGO, or entities that are specific for a certain organization. 
Besides offering a vocabulary, KGs show the relationships between entities, uncovering more complex interrelations between them.
KGs are typically organized in triples inspired by natural language containing subject, predicate and object. Wide-spread models for constructing and interchanging KGs are the Resource Description Framework (RDF) and JSON-LD.
Due to their capabilities, we take knowledge graphs as base technology for enterprise modeling in the following elaborations.


While enterprise modeling is a resource and time-intensive task, which has originally been conceived for human actors, knowledge graphs can enable the semantic processing of the created models by machines through adding a formal semantic layer to them. Thereby, semantic information is either added to the models ex-post, also denoted as \emph{semantic annotation} or \emph{semantic lifting}~\cite{Fill17,Fill18,Fill11}, or, knowledge graphs are used a-priori as an input source for the automated generation of enterprise models~\cite{9233131}.

As shown in~\cite{reitemeyer2019ontology}, semantic annotations can for example be added while creating BPMN models. Thereby knowledge graph nodes describe the referring model element in terms of the modeling language as well as the content representing the type semantics and inherent semantics of the model element.
Another example is the approach by Smajevic and Bork where they use knowledge graphs for ArchiMate to detect enterprise architecture smells~\cite{smajevic2021using}. 
The approach is based on an ex-post transformation of an ArchiMate model into a knowledge graph, which is then input for the enterprise architecture smell detection.

Further examples of the application of semantic technologies in enterprise modeling include the usage of requirements ontologies, which are employed for rule-based mapping to BPMN elements and subsequent generation of a BPMN graph~\cite{yanuarifiani2019automating}. Additionally, formal knowledge bases containing domain knowledge are utilized for the generation and annotation of business process models~\cite{riehle2017automatically}.

\subsection{Large Language Models and Semantics}
Since the release of ChatGPT in 2022, Generative Pre-Trained Transformer (GPT) models have been widely adopted for various use cases. Especially ChatGPT, with its specialization on conversations and artificial intelligence based responses to user input~\cite{campbell2020beyond} has been applied in a wide range of private and business scenarios, leading quickly to over 100 million active users~\cite{chow2023chatgpt}. 

GPTs are based on Large Language Models (LLM), which have also found wide acceptance in use cases such as image recognition, speech-to-text or text processing tasks~\cite{vaswani2017attention,xu2023multimodal}. In general, LLMs make experimentation with Artificial Intelligence more accessible 
due to their capability of specifying natural language prompts for triggering generations of text or images~\cite{wu2022promptchainer}.
They thus seem optimally suited for tasks in language-based conceptual enterprise modeling.

While traditional approaches in semantic mapping are mature,  first approaches for LLM-based semantic mappings have emerged.
For example, Wang et al. \cite{wang2023exploring} developed an approach based on LLMs for biomedical concept linking. Their in-context approach follows a two stage procedure: at first the biomedical concepts are embedded into the overall context via a prompt and then similarity mapping is performed to get top candidates to match with an input concept.

In another use case, Hertling and Paulheim~\cite{hertling2023olala} developed an approach for concept matching in knowledge graphs. It addresses the problem that real-world objects may be contained in multiple knowledge graphs and one wants to determine whether two objects are equivalent. Their approach uses open source LLMs to match candidate concepts from two different knowledge graph inputs using cardinality and confidence filters to improve result quality. As stated by the authors, the approach outperforms comparable approaches even though it is only based on natural language descriptions.
They argue that semantics in knowledge graphs are typically described with either natural language in labels, comments or descriptions, relations in between concepts, or formal axioms. While in the past, the natural language semantics were only targeted at humans, LLMs now add powerful machine-processing capabilities for these natural language descriptions. 
The results so far showed that LLMs can lead to improvements in enterprise modeling as well as in semantic concept mapping. Therefore, we will show in the next chapter how LLMs can be used in concept mapping for automated enterprise modeling. This will permit to evaluate which model element is most similar to a given real-world concept and to show the underlying explanation via the LLM.

\subsection{Enterprise Modeling and Large Language Models}

Recently, the application of LLMs in enterprise modeling has been explored. For example, Fill et al.~\cite{fill2023conceptual} conducted experiments for investigating the capabilities of LLMs in the creation and interpretation of models in different enterprise contexts such as business process, systems, and data modeling. The authors concluded, that LLMs showed a huge potential for supporting modeling tasks with potential for improvement especially in terms of evaluation, in finding the right modeling languages and notations to use with LLMs, and in regard to the trade-offs between open source LLMs versus commercial ones.

Further on, Härer~\cite{harer2023conceptual} designed an architecture for generating PlantUML and Graphviz models based on LLMs in a conversational style. His study aimed at implementing a conceptual model interpreter for LLMs focussing on generating models with the correct syntax. He concludes that iterative modeling using GPT-4 is generally possible in a conversational fashion.

Vidgof et al.~\cite{vidgof2023large} discuss the usage of LLMs in the Business Process Management lifecycle. They suggest the usage of LLMs in explaining business process models as a model chatbot to answer queries a user may have about a concrete model, or as process orchestrator.

Barn et al.~\cite{BarnBS23} investigate the adaptations that have to be made in enterprise modeling languages for enabling prompt-based interactions. They develop a prompt engineering meta-model including domain concepts as well as modeling language elements for the 4EM method.

With a focus on software modeling, Camara et al.~\cite{camara2023assessment} investigated the capabilities of ChatGPT in modeling UML by generating PlantUML code. They find that ChatGPT-based software modeling has limitations in terms of syntax, semantics, consistency, and scalability, especially when compared to code generation.

For better addressing the semantics of modeling languages, it is also being explored how knowledge graphs can be employed. Approaches in this regard include three areas of research with a high potential for further evaluation, namely: (1) \emph{Knowledge Graph-enhanced LLMs}, for improving the knowledge of LLMs during the pre-training and the inference phase, (2) \emph{LLM-augmented Knowledge Graphs}, including LLMs for various tasks such as graph construction or question answering, and (3) \emph{Synergized LLMs + Knowledge Graphs}, for bidirectional enhancement of knowledge graphs and LLMs~\cite{pan2024unifying}.
Luo et al.~\cite{luo2023reasoning} argue that LLMs are skilled in reasoning in complex tasks, but struggle with up-to-date knowledge. Additional, they can lack from hallucinations in reasoning leading to negative impact in terms of performance and trustworthiness. Therefore, they developed the method ~\emph{reasoning on graphs} to enable faithful and interpretable reasoning in LLMs.
Finding the right balance between performance and efficiency is a key task for knowledge-based systems. LLMs could help to improve the performance for example in solving some of the knowledge-intensive sub tasks such as mention detection, entity disambiguation, or relation detection~\cite{hu2023empirical}.

These early results show that knowledge graphs and LLMs can support each other in both directions. The use of knowledge graphs in the context of enterprise modeling could also have an impact on LLMs.

\section{Integration of LLMs in Enterprise Modeling}
\label{sec:integration}
\begin{figure}
\begin{center}
    \includegraphics[width=250pt]{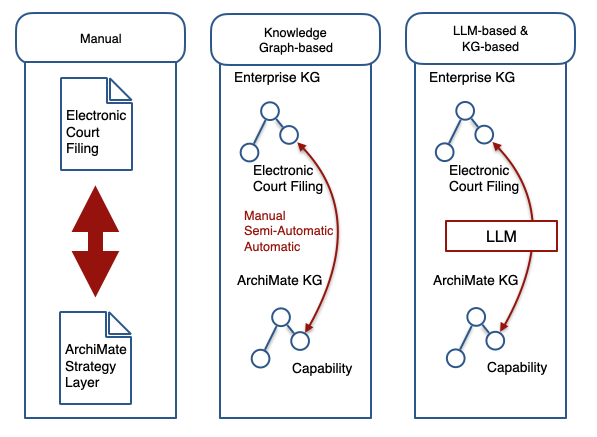}
\caption{Options for semantic mapping between the ArchiMate element capability and the domain concept Electronic Court Filing.} \label{fig4}
\end{center}
\end{figure}

In the following, three options for relating domain concepts with modeling language concepts will be explored: \emph{manual}, \emph{knowledge graph-based}, and \emph{LLM and knowledge graph-based}. All options are described with reference to the exemplary domain concept of \emph{Electronic Court Filing} as part of the U.S. National Information Exchange Model (NIEM), which provides an open vocabulary for exchanging information between public and private organizations\footnote{See https://www.niem.gov/}. Accordingly, the domain concept is to be mapped to the ArchiMate element designated \emph{Capability}. For all three options, we provide a description of the input, including the baseline information for the aforementioned concepts, the processing steps necessary for identifying a suitable relation, and the resulting output, as illustrated in Figure \ref{fig4}.

A more detailed description of the mapping example will be provided in Section \ref{sec:evaluation}, as it forms part of one of the use cases employed in the evaluation.

\subsection{Manual}
As previously outlined in Section 2, one of the illustrative examples is the modelling of an ArchiMate capability map viewpoint. A capability map is a tool that is typically employed to gain a structured overview of an enterprise’s capabilities. In the aforementioned viewpoint, three ArchiMate elements may be utilized: \emph{Outcome}, \emph{Capability}, and \emph{Resource}. For example, a law firm may be confronted with evaluating the services offered in a service repository, such as NIEM for the justice domain. One of the service elements is \emph{Electronic Court Filing}. An enterprise architect is consulted for the task and has to decide based on the natural language descriptions of the ArchiMate elements and the domain element \emph{Electronic Court Filing} how to model it. 

In NIEM, the term \emph{Electronic Court Filing} is defined as: "The LegalXML Electronic Court Filing 5.0 (ECF 5.0) specification consists of a set of non-proprietary XML and Web services specifications along with clarifying explanations and amendments to those specifications that have been added for the purpose of promoting interoperability among electronic court filing vendors and systems.  ECF Version 5.0 is a major release and brings the specification into conformance with the National Information Exchange Model (NIEM) 4.0."\footnote{https://www.niem.gov/about-niem/iepd-registry/electronic-court-filing-version-50}. 

In the ArchiMate standard specification, element descriptions contain two parts: a \emph{short} description and a more \emph{comprehensive} one. The short description offers a brief overview of the concept, whereas the comprehensive description refers to the standards set forth in the ArchiMate standard. For the concept \emph{Capability}, the short description is: "A capability represents an ability that an active structure element, such as an organization, person, or system, possesses."\footnote{https://pubs.opengroup.org/architecture/archimate3-doc/ch-Strategy-Layer.html}. The two parts of the description are employed as inputs for the decision regarding instantiation.

The selection of the most appropriate ArchiMate element for a specific domain concept, such as Electronic Court Filing, is typically based on a combination of factors, including the  descriptions, modeling language and business rules, experience, and the ability to address a stakeholder concern. 

The modeling process, which considers the domain, model elements, and further context, results in the instantiation of a model element. For example, the \emph{Electronic Court Filing} is an instantiated ArchiMate \emph{Capability} as illustrated on the left side of Figure \ref{fig1}.

\subsection{Knowledge Graph-based}
The use of knowledge graphs (KGs) as a foundation for determining the instantiation of a domain concept as a modeling language element initially entails the adaptation of the format of the processed input. In the manual approach, the input descriptions of the domain element and ArchiMate elements were in the form of natural language. In contrast, in the KG-based approach, machine-processable formats are employed. For our example, an ArchiMate knowledge graph is employed, encompassing the ArchiMate modeling language, its constituent concepts, their interrelationships, and associated application rules. This is integrated with a NIEM enterprise KG, which includes the NIEM concepts.

\begin{figure}
\begin{center}
    \includegraphics[width=350pt]{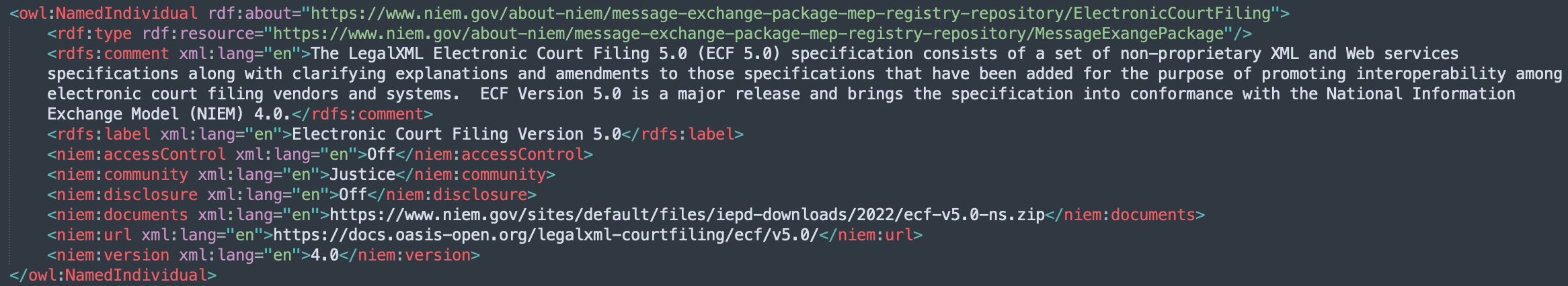}
\caption{Example for Electronic Court Filing and its properties as knowledge graph entities.} \label{fig8}
\end{center}
\end{figure}

A second alteration concerning the manual approach pertains to the processing of the input data. Notwithstanding the fact that the input format is designed to facilitate machine processing, it can also be processed manually by human experts. This entails taking the KG elements and performing the instantiation process as described in section 3.1.
The principal advantage of machine-processable data formats is the option of automatic, machine-based processing. With regard to knowledge graphs (KGs), the integration of two concepts is frequently described as a semantic mapping process based on approaches for elaborating semantic similarity. This process was initially focused on linking metadata from hypermedia systems~\cite{carr2001conceptual} or linking subsections of online textbooks by using domain knowledge (such as other textbooks or educational resources)~\cite{meng2016knowledge}. The concept of semantic similarity is further developed through the introduction of quantitative values to the relationship between two concepts in a knowledge graph (KG). This enables not only the identification of related concepts, but also the evaluation of the strength of these relationships in terms of the structure and information content of the concepts themselves~\cite{zhu2016computing}.\\
In corpus-based approaches, semantic similarity between concepts is gauged using data obtained from extensive corpora, such as Wikipedia. These approaches are based on semantic relatedness, rather than semantic similarity, and thus do not take into account hierarchical relations~\cite{turney2010frequency}. In contrast, knowledge-based approaches assess semantic similarity by incorporating the hierarchy of concepts in the graph, such as the least common subsumer of two concepts. Additionally, approaches like Wpath integrate corpus-based and knowledge-based information for semantic mapping~\cite{zhu2016computing}.
Thirdly, semi-automatic processing approaches are feasible, utilizing automated semantic similarity measures and human expert experience to facilitate mapping tasks.

The output of all of these approaches in KG-based integration, whether manual, semi-automated, or automated, is the initialised modeling element. Additionally, it comprises the formalized relation between the domain element and the ArchiMate element. In the case of the aforementioned statistical approaches, it also includes a measure of the similarity.

\subsection{LLM-based and KG-based}
Next, the combination of LLM-based and KG-based approaches shall be discussed. As outlined in section 3.2., they employ knowledge graphs as an input. The most significant alteration is the shift in processing methodology, moving away from semantic similarity approaches and towards the utilisation of LLMs for the assessment of domain concept instantiation within a modeling language. This also leads to the generation of the instantiated model element and the KG relation between the elements. In contrast to KG-based approaches, which typically produce the same output format as the input, LLM-based approaches offer the option of generating formats such as JSON for subsequent processing. The use of KG-based inputs guarantees that the LLM processes curated and reliable knowledge sources, thereby ensuring the independence of the results from the training of the LLM. In terms of processing, the KG-based approach employs statistical measures based on the relation of the concepts in the KG, whereas LLM-based approaches utilize language-based probability. This results in altered outcomes, as LLM-based approaches do not incorporate a definitive measure of the relatedness between two elements; instead, they merely delineate it in natural language, as illustrated in Table \ref{tab:relationTypes}. While KG-based approaches may necessitate preliminary processing to facilitate integration between two KGs and, thus, enable statistical calculations, LLM-based processing does not require such steps, as it is capable of processing KG formats.

\begin{table}
\includegraphics[width=280pt]{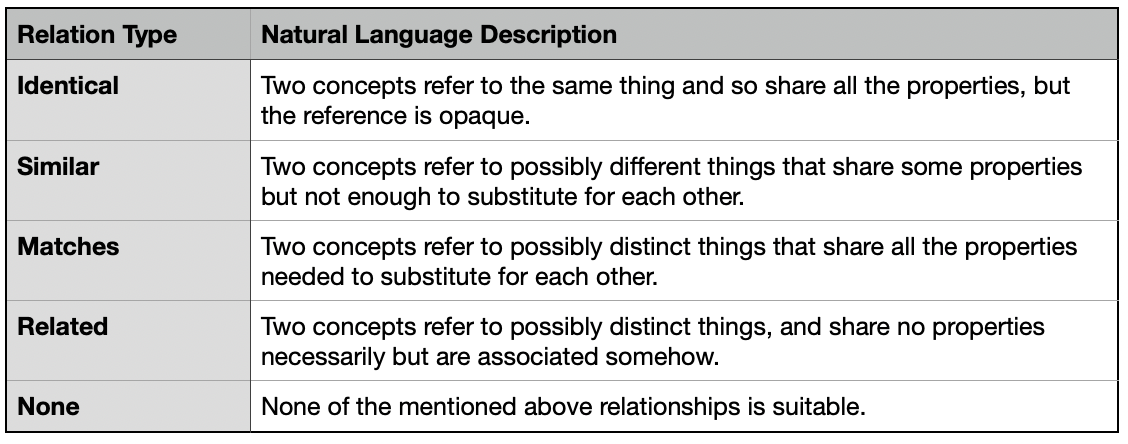}
\caption{Relation types for defining relationships between knowledge graph elements (based on~\cite{halpin2010owl}).}
\label{tab:relationTypes}
\end{table}

\section{Empirical Evaluation}
\label{sec:evaluation}
In order to examine the ability of large language models to instantiate a domain concept within a modeling language, we employed a two-pronged approach based on our previous work in this area~\cite{reitemeyer2024leveraging}. First, an online survey was conducted with experts from various domains to establish a baseline for how human actors would complete the  tasks. Subsequently, a series of experiments was conducted with ChatGPT-4o. In both cases, domain concepts were observed in regard to ArchiMate elements and ranked according to a set of pre-defined criteria. The results were then investigated in terms of their proximity and relation types.
This section outlines the general context of the investigations, followed by a detailed account of the survey and experimental procedures.

In the design of the general setting, five real-world use cases were selected for the framing of tasks. These cases were all situated within the domain of justice. For each of these use cases, one specific domain concept (Electronic Court Filing, Case Opening and Docketing, Court File, Court and Administration Mailbox, Participants)  was identified as a requisite element for instantiation within an ArchiMate model. The model elements were constrained by the utilization of particular ArchiMate viewpoints, each of which sought to address a specific stakeholder concern and included a limited set of model elements.

For both approaches, the input was structured with a brief description of the functional context, a concrete description of the domain concept, the relevant ArchiMate viewpoint, and its associated model elements.

Finally, the tasks were articulated: 1) Ranking of the ArchiMate elements in terms of their suitability to the domain concepts, from best to worst; 2) defining how close the relation between a domain concept and the ArchiMate concepts are considered, based on five values from \emph{very high} to \emph{very low}; 3) defining the relation type between domain concept and ArchiMate elements based on the relation types \emph{identical}, \emph{similar}, \emph{matches}, \emph{related} or \emph{none} as defined in Table \ref{tab:relationTypes}.

\subsection{Expert Assessment}
The survey was designed with the aforementioned considerations about the general setting in mind. The survey was constructed using LimeSurvey\footnote[1]{https://www.limesurvey.org/de}. In the initial phase of the survey, a pre-test of the questionnaire was conducted, which served as the foundation for subsequent internal feedback and reviews. Following the incorporation of these insights, the final survey was constructed~\cite{kallus2016erstellung}.
The participants were selected from academia, public and private industry~\cite{kirchhoff2010fragebogen}. Twelve experts responded to the survey. As illustrated in Fig. 4, their ArchiMate expertise ranges from first experience to experienced master.

\begin{figure}[ht]
\includegraphics[width=\textwidth]{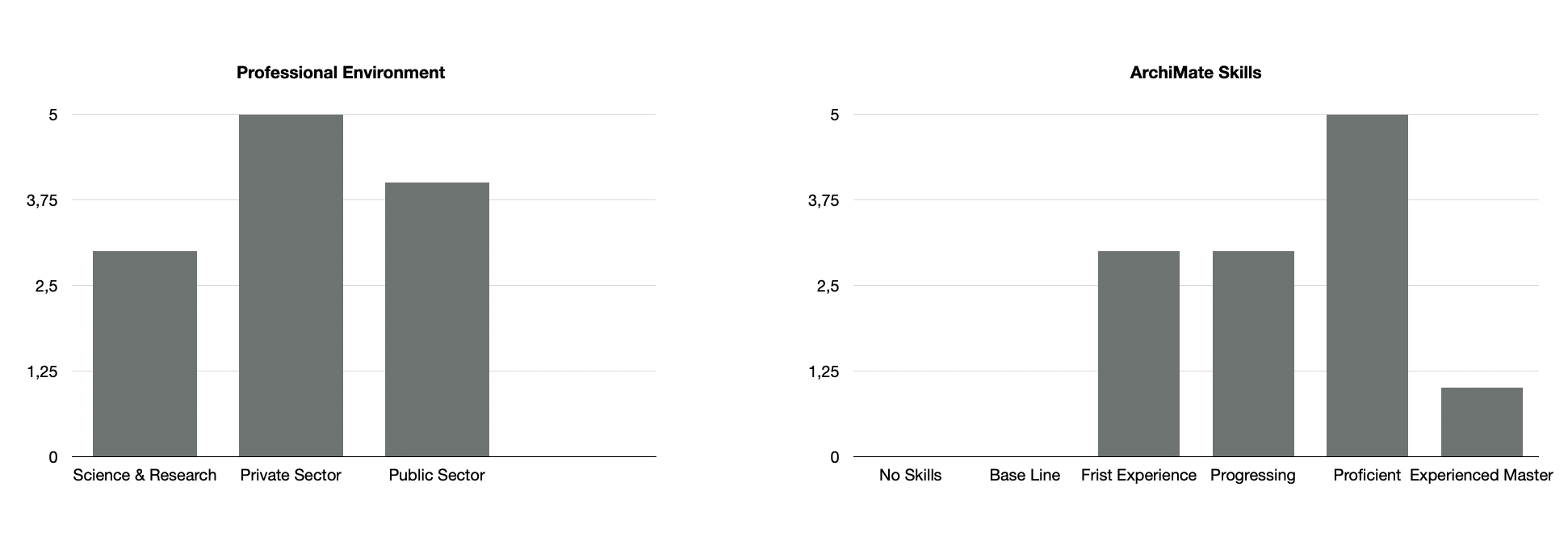}
\caption{Professional background and ArchiMate skills of participants.} \label{fig5}
\end{figure}
A preliminary data analysis and evaluation~\cite{raab2015fragebogen} was conducted using the internal tools of LimeSurvey and subsequently assessed.
With regard to RQ1 ('prioritization of ArchiMate elements'), the experts demonstrated a high degree of clarity regarding the first two positions in the ranking. The percentage of experts ranged from 75\% in Case 2 to 50\% in Case 4 for the first rank and from 58\% in Case 5 to 33\%  in Case 1 for the second rank. As the chosen ArchiMate viewpoints contained between three and five possible elements and the option to only include relevant elements in the ranking, a meaningful comparison of further ranks was not feasible. The option to refrain from selecting each and every element resulted in a range of 18\% in Case 5 to 10\% in Case 4 of unused elements in the rankings.

In regard to RQ2 ('probability of proposing an instance of an ArchiMate element'), the experts indicated a preference for the \emph{Very High}, \emph{High} and \emph{Very Low} categories when asked to assess the likelihood of any of the potential viewpoint elements being instantiated. Conversely, the \emph{Low} and \emph{Medium} categories were perceived as less probable, although there was a degree of variation across all possible values.
Figure \ref{fig:rq2rq3} depicts the findings of RQ2 and RQ3, which sought to identify the optimal relationship between the ArchiMate elements of the viewpoints and the domain concept as outlined in Tab. 1. The most preferred relation was \emph{Related}, followed by \emph{Similar} and \emph{None}. 

\begin{figure}[ht]
\includegraphics[width=\textwidth]{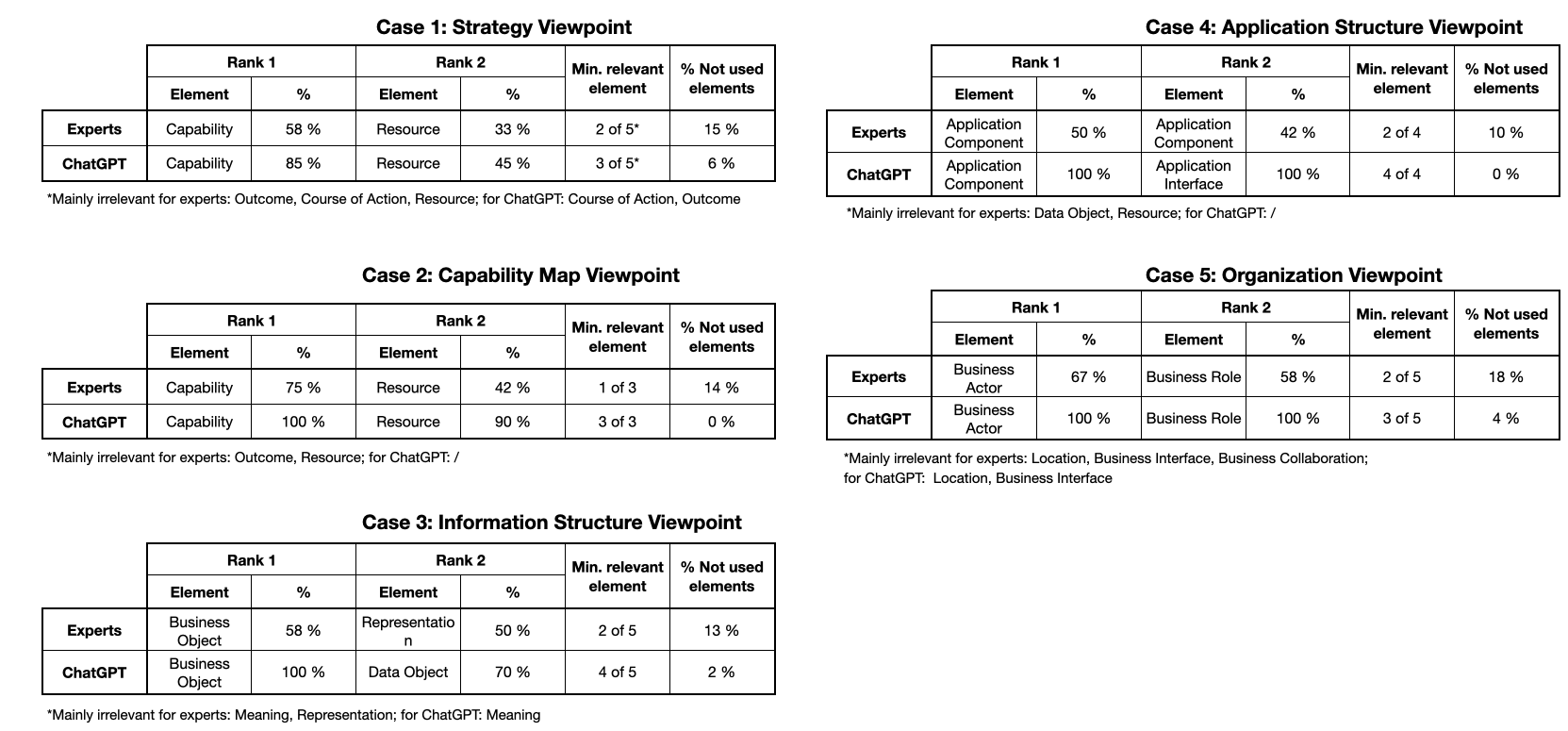}
\caption{Results of RQ1 showing the best two elements in the ranking and the quota of irrelevant elements.} \label{fig:rq1}
\end{figure}

\subsection{LLM-based Assessment}
To ensure the consistency of the results, the knowledge graph-based experiments were conducted 20 times on ChatGPT-4o.
The prompts were designed based on the aforementioned structure, but with the distinction that the context description of domain, ArchiMate, and relation types were made accessible for the prompts through the use of knowledge graphs.

Subsequently, the optimal prompting technique was explored. Three techniques were deemed feasible: 1) \emph{zero-shot prompting} as a technique in which the task is based on natural language and entered in a single description at the time of inference. In this approach, no examples are provided~\cite{amini2024towards}.
2) In \emph{few-shot prompting} approaches, task examples are provided, including context and results, which support the LLM in its understanding~\cite{brown2020language}.
3) \emph{Chain-of-thought} prompting, in which examples of the underlying thought process are provided, guiding the model to a series of reasoning steps necessary to reach the result~\cite{wei2022chain}. To be consistent with the expert study, zero-shot prompting was chosen.

For each of the five use cases, the initial prompt delineated the context and posed the question for RQ1. Subsequently, prompts were presented for each of the questions for RQ2 and RQ3. To emulate the insights gleaned from the survey, all five use cases were conducted within the same chat.
\begin{figure}
\includegraphics[width=\textwidth]{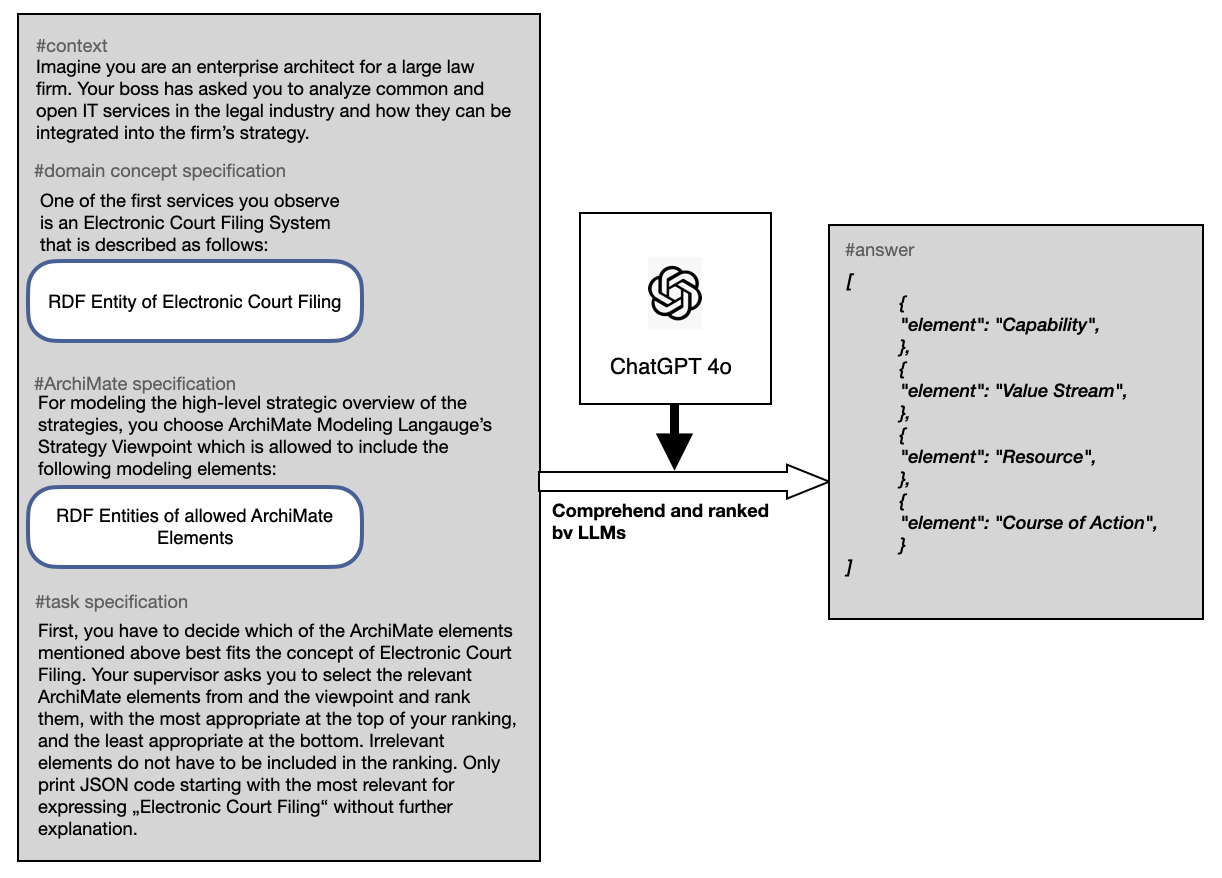}
\caption{Experiment prompt for RQ1 consisting of context, domain concept specification, ArchiMate specification, and task specification.} \label{fig:prompt}
\end{figure}

In regard to RQ1, ChatGPT consistently demonstrated a high degree of consistency in its selections, with a percentage between 100\% and 85\% consensus for the first rank, and between 100\% and 45\% percent consensus for the second rank. In two cases, all of the ArchiMate elements were deemed relevant, while in no more than 6\% of the cases were elements deemed irrelevant.

\begin{figure}
\begin{center}
\includegraphics[width=0.8\textwidth]{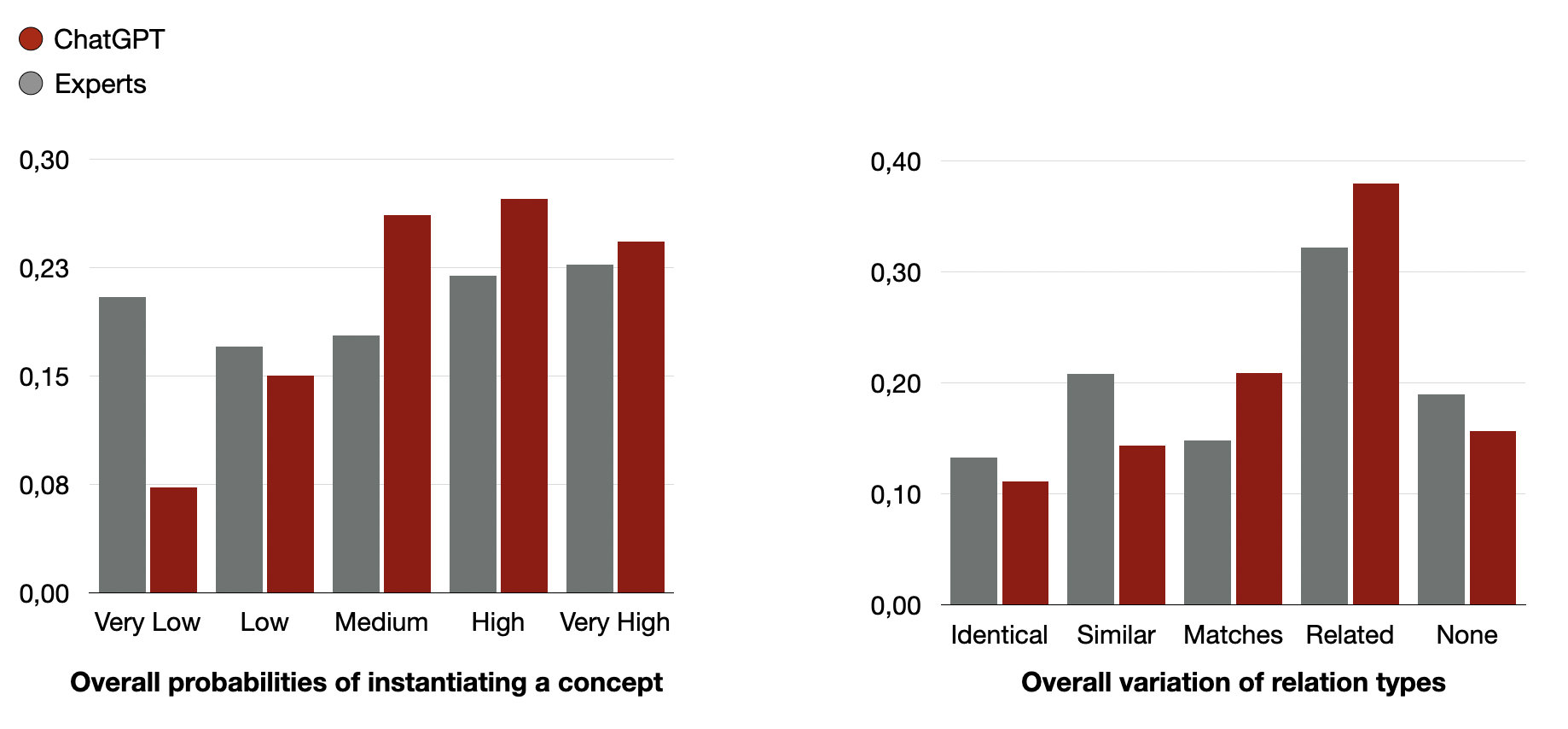}
\caption{Results of RQ2 und RQ3 showing the overall probabilities of instantiating a concept and the variation of used relation types between domain elements and ArchiMate elements.} \label{fig:rq2rq3}
\end{center}
\end{figure}

In regard to RQ2, the ChatGPT actors exhibited a proclivity for selecting the values \emph{Very High}, \emph{High}, and \emph{Medium}. Conversely, they demonstrated a diminished propensity for selecting \emph{Low} and \emph{Very Low} probabilities, which were less frequently instantiated as ArchiMate elements. With regard to RQ3, the ChatGPT actors predominantly selected \emph{Related} and \emph{Matches}, while \emph{Similar} and \emph{None} were less frequent. \emph{Identical} was the least frequent selection.

\section{Discussion}
\label{sec:discussion}

In general, the results demonstrated a certain degree of inconsistency and ambiguity in the experts' opinions regarding the selection of the appropriate element for instantiation. ChatGPT exhibited a higher degree of accuracy, but often considered all elements to be relevant. In general, both experts and the LLM identified the same element as the most relevant. This demonstrates that LLMs are capable of selecting an appropriate element with minimal variation. However, they exhibit a limitation in discerning irrelevant elements, which could potentially result in erroneous instantiation in more intricate domain descriptions. For instance, ChatGPT selected 25\% of the irrelevant elements with a \emph{High} probability of instantiation.

The results for RQ2 demonstrated a notable divergence from the judgments of human experts, particularly with regard to the assignment of \emph{Low} and \emph{Very Low} probabilities for the instantiation of ArchiMate elements. In light of the aforementioned inconsistency of ChatGPT with regard to irrelevant elements, it demonstrated remarkable consistency in mapping probabilities to specified ranks. All elements on rank 1 were specified to be \emph{Very High}, while experts had just 75\%. For rank 2 it was still 95\% for ChatGPT and 57\% for experts, while medium was 86\% for ChatGPT and 41\% for experts.

The experts and ChatGPT encountered difficulty with RQ3, where the expected relation types should have been '\emph{Related}' or '\emph{None}'. '\emph{Identical}' requires two concepts to be the exact same thing, which was not given in this context. '\emph{Similar}' demands at least some shared properties, while '\emph{Matches}' refers to the same properties that can be substitutes for each other. The results indicate that experts selected narrower relations, such as '\emph{Identical}' and '\emph{Similar}', while ChatGPT favored lose relation types, such as '\emph{Matches}' and '\emph{Related}'.

As the relation types and their definitions are based on an ontological context~\cite{halpin2010owl}, including formal definitions, the descriptions may lack sufficient clarity to be understood in an appropriate manner. Furthermore, a more comprehensive inquiry into the characteristics of relations, including their directionality, would facilitate the enhancement of the knowledge graph base. The results of the experiment demonstrate that, while ChatGPT performs the tasks with less variability than human experts, caution should be exercised when interpreting the results.

The remaining variability and inconsistencies let us conclude that, despite the precision with which LLMs can rank and instantiate, relying on them alone for modeling without the input of human modelers may be inadvisable, in particular as the use cases were simplified for the experiments in comparison to real world scenarios. In conclusion, novel modeling approaches that integrate the strengths of LLMs in processing comprehensive data and proposing drafts of models, while leveraging human expertise for ensuring semantic correctness, are necessary. However, the challenge remains that a modeler must ultimately address the complexity of the utilized input sources.
We can thus identify a number of limitations to inform future research. First, only structure-oriented viewpoints were selected so far, while flow-oriented viewpoints were excluded. Secondly, the meaningfulness of the results may be enhanced through a larger number of human expert participants. Thirdly, the discussion on reliability in more complex environments, as well as the relationship between modeling and domain concepts, could be more fully addressed through the use of more complex and detailed use cases.

\section{Conclusion and Outlook}
\label{sec:conclusion}
This paper examined the challenges and opportunities of using large language models in enterprise modeling. An experiment was conducted in which domain concepts and ArchiMate model elements as knowledge graphs were used as input for LLM-based modeling. As baseline, an expert survey was conducted. The results show that, although ChatGPT exhibited greater consistency than human experts, it still demonstrated variability and inconsistency. This indicates that human modelers are still necessary to validate the machine-created results. The conclusion and limitations indicate two areas for future research: 1) advanced experiments on the capabilities of LLMs in modeling, including more detailed use cases, and 2) developing a modeling process and demonstrating its conceptual and technical feasibility, integrating LLM capabilities with human expertise.

\newpage

%
%
\bibliographystyle{splncs04}
\bibliography{bibliography}
\end{document}